\documentclass[aps,11pt]{revtex4}
\usepackage[dvips]{graphicx}
\begin{document}

\title{Crystal-field ground state for the localized $f$ state\\
in heavy fermion metal YbRh$_{2}$Si$_{2}$$^{+}$}
\author{R. J. Radwanski}
\homepage{http://www.css-physics.edu.pl} \email{sfradwan@cyf-kr.edu.pl}
\affiliation{Center of Solid State Physics,
S$^{nt}$ Filip 5, 31-150 Krakow, Poland,\\
@ Institute of Physics, Pedagogical University, 30-084 Krakow, Poland}
\author{Z. Ropka}
\affiliation{Center of Solid State Physics, S$^{nt}$ Filip 5, 31-150 Krakow, Poland}
\begin{abstract}
We fully support the interpretation of Prof. F. Steglich, reported recently in Phys. Rev. Lett. \textbf{91} (2003)
156401, that the observed \textit{below} Kondo temperature $T_K$, Electron Spin Resonance (ESR) in a heavy fermion
metal YbRh$_{2}$Si$_{2}$ is associated with the Yb$^{3+}$ moment, more exactly with the strongly-correlated 4$f^{13}$
configuration. We derived the ground-state eigenfunction $\Gamma _{7}^{1}$ = $0.803 |\pm 3/2>$ + 0.595 $|\mp 5/2>$ -
0.026 $|\mp 1/2>$ - 0.008 $|\pm 7/2>$ and crystal field parameters: $B_{2}^{0}$ = +14 K, $B_{4}^{0}$ = +60 mK,
$B_{6}^{0}$ = -0.5 mK, $B_{4}^{4}$ = -2.23 K and $B_{6}^{4}$ = -10 mK with a small orthorhombic distortion $B_{2}^{2}$
= +0.23 K, that perfectly reproduce the observed ESR values of $g_\bot $= 3.561 and $g_\| $ = 0.17. These parameters
also describe the observed overall temperature dependence of the paramagnetic susceptibility and predict anomalous
temperature dependence of quadrupolar interactions.

PACS: 71.27.+a, 75.20.Hr, 76.30.-v

Keywords: heavy fermion metal, crystal field, YbRh$_{2}$Si$_{2}$
\end{abstract}
\maketitle

Recently Sichelschmidt \textit{et al.} \cite{1} reported the first successful Electron Spin Resonance (ESR) studies on
single crystalline heavy fermion metallic compound YbRh$_{2}$Si$_{2}$ that allowed for the unambiguous observation of
the localized $f$ states in a Kondo compound. The importance of this experiment relies in the fact, that practically
all theories devoted to heavy-fermion phenomena were taking the itinerant or band behavior of $f$ electron as the
starting undoubt point. Theoretical approaches to heavy-fermion phenomena with localized $f$ states were simply
rejected both by referees of prestigious physical journals and by their editors that largely prohibited any discussion
of the localized magnetism and crystal-field interactions. Thus, the discovery by the Prof. F. Steglich's group is
really of the great importance for theoretical understanding of heavy-fermion compounds. The authors of Ref. \cite{1}
are fully aware of the theoretical importance of their observation putting a lot of attention to evidence that this ESR
signal comes out from the bulk unscreened $Yb^{3+}$ moments \textit{below} Kondo temperature $T_K$. By means of the ESR
experiment, at B=0.188 T with the X-band frequency of 9.4 GHz, Sichelschmidt \textit{et al.} \cite{1} revealed the
existence of a localized doublet characterized by very anisotropic $g$ tensor with $g_\bot$ = 3.561 and $g_\|$ =0.17.
The authors of Ref. \cite{1} have attributed this doublet to a Kramers doublet related to the Yb$^{3+}$ configuration.
We fully agree with this interpretation.

The aim of this short paper is to report our search for the atomic-scale CEF parameters describing the observed $g$
tensor in order to find the microscopic origin of this localized state.

We have attributed, as in Ref. \cite{1}, the observed state to the Yb$^{3+}$ configuration, more exactly to the
strongly-correlated 4$f^{13}$ configuration, in the Hund's rule ground multiplet $^{2}F_{7/2}$ described by $J$ = 7/2 -
our studies show that the higher multiplet $^{2}F_{5/2}$ does not practically affect the ground-multiplet properties.
The crystal field of the tetragonal symmetry:

$H_{CF}^{tetr} $ = $B_{2}^{0}O_{2}^{0}$ + $B_{4}^{0}O_{4}^{0} $ + $B_{6}^{0}O_{6}^{0}$ + $B_{4}^{4}O_{4}^{4}$ +
$B_{6}^{4}O_{6}^{4}$

splits this 8-fold degenerated multiplet into 4 Kramers doublets, 2$\Gamma _{6}$ and 2$\Gamma _{7}$. The lower in
energy $\Gamma _{6}$ state takes the form (notation by $\sin \alpha$ and $\cos \alpha$ assures the automatic
normalization and the sign $\pm$ corresponds to 2 Kramers conjugate states):

$\Gamma _{6}^{1}$ = $\sin \alpha |\pm 1/2>$ + $\cos \alpha |\mp 7/2>$ whereas

$\Gamma _{6}^{2}$ = $\sin \alpha |\pm 1/2>$ - $\cos \alpha |\mp 7/2>$ is higher in the energy.

Similarly, the lower energy $\Gamma _{7}$ state takes the form:

$\Gamma _{7}^{1}$ = $\sin \beta |\pm 3/2>$ + $\cos \beta |\mp 5/2>$ and the higher one:

$\Gamma _{7}^{2}$ = $\sin \beta |\pm 3/2>$ - $\cos \beta |\mp 5/2>$.

Thus as the ground state the only states $\Gamma _{6}^{1}$ and $\Gamma _{7}^{1}$ can be.

We intend to reproduce the $g$ tensor keeping the overall CEF splitting $\Delta _{CEF} $ of size of 600-750 K (55-70
meV) and the first excited level $D$ at 70-100 K (6-9 meV) in order to assure the proper thermodynamics and a
reasonable magnitude of CEF interactions.

\textbf{$\Gamma _{6}^{1}$ ground state}

$\Gamma _{6}^{1}$ = 0.883 $|\pm 1/2>$ + 0.469 $|\mp 7/2>$, with $\alpha$ = 62.0 is characterized by $J_\bot $ =$\pm
1.56$ (with $g_L$ = 8/7 it gives $m_{x}$ = $\pm 1.78$ and $g_\bot$ = 3.56 like in the ESR experiment), $J_\|$ = $\pm
0.38$ that is much larger than in experiment $\pm 0.08$. The quadrupolar operator value $Q_f$ = $3J_{z}^{2}$ - J(J+1)
of -7.1.

This state can be obtained as the ground state by tetragonal CEF interactions $B_{2}^{0}$ = +6 K, $B_{4}^{0}$ = -120
mK, $B_{6}^{0}$ = +8 mK, and $B_{4}^{4}$ = -3.032 K, for instance. For these parameters $\Delta _{CEF} $ = 700 K and
$D$=64 K.

\textbf{$\Gamma _{7}^{1}$ ground state}

The value of $J_\bot$ = $\pm 1.56$ resulting from the ESR experiment $g_\bot$ =3.561 provides the ground-state
eigenfunction:

$\Gamma _{7}^{1}$ = 0.847 $|\pm 3/2>$ + 0.532 $|\mp 5/2>$, with $\beta$ = 57.89.

It yields $J_\|$ = $\pm 0.36$ and $Q_f$ = -5.6. A value of $J_\|$ is much larger than in experiment ($\pm 0.08$).

This state can be obtained as the ground state by tetragonal CEF interactions: $B_{2}^{0}$= +18 K, $B_{4}^{0}$ = +60
mK, $B_{6}^{0}$ = -2 mK, and $B_{4}^{4}$ = -2.147 K, for instance. These parameters yield the overall CEF splitting
$\Delta _{CEF}$ = 770 K and the first excited doublet at D=75 K.

If we allow for a departure from the strict reproduction of $J_\bot$ = $\pm 1.56$ in order to make better agreement for
$J_\|$ then we could come to parameters, for instance, $B_{2}^{0}$ = +14 K, $B_{4}^{0}$ = +60 mK, $B_{6}^{0}$ = -0.5
mK, $B_{4}^{4}$ = -1.95 K and $B_{6}^{4}$ = -10 mK. These parameters yield: $\Gamma _{7}^{1}$ = 0.817 $|\pm 3/2>$ +
0.576 $|\mp 5/2>$, with $\beta$ = 54.8 - it is characterized by $J_\bot$ = $\pm 1.63$, $J_\|$ = $\pm 0.18$ and $Q_f$ of
-6.0. By a 4.5 percent increase of $J_\bot$ we get 50 percent decrease of $J_\|$ keeping the tetragonal symmetry.

The perfect reproduction of the ESR results, $g_\bot$= 3.561 (== $J_\bot$ = $\pm 1.56$ and $g_\|$ = 0.17 (== $J_\|$ =
$\pm 0.08$) is obtained for parameters: $B_{2}^{0}$= +14 K, $B_{4}^{0}$ = +60 mK, $B_{6}^{0}$ = -0.5 mK, $B_{4}^{4}$ =
-2.23 K and $B_{6}^{4}$=-10 mK with a small local orthorhombic distortion $B_{2}^{2}$ = +0.23 K. These parameters
yield:

$\Gamma _{7}^{1}$ = 0.803 $|\pm 3/2>$ + 0.595 $|\mp 5/2>$ - 0.026 $|\mp 1/2>$ - 0.008 $|\pm 7/2>$

with $J_\bot$ = $\pm 1.565$, $J_\|$ = $\pm 0.081$ and $Q_f$ = -4.7. The excited states are at 93 K ($\Gamma _{6}^{1}$
with $Q_f $ = -13.8), 485 K ($\Gamma _{7}^{2}$) and 680 K ($\Gamma _{6}^{2}$). It is important to realize that whatever
lower symmetry is only 4 Kramers doublets always are for the 4$f^{13}$ configuration.

This electronic structure is very plausible, though we do not think that it is the final one. For it thermodynamical
properties have to be more carefully analyzed, the best would be the direct inelastic-neutron-scattering experiment.
However, apart of the g tensor these parameters reproduce the overall temperature dependence of the paramagnetic
susceptibility $\chi (T)$ and its huge anisotropy, presented in Fig. 1a of Ref. \cite{2}, the preference for the
magnetic ordering with moments perpendicular to the c axis, the magnetization curve for external magnetic fields up to
60 T applied along the tetragonal c axis (Fig. 1b of Ref. \cite{2} - the magnetization at 2 K and at 60 T amounts to
0.85 $\mu _{B}$). The derived electronic structure predicts a Schottky-type contribution to the specific heat with the
maximum of 3.62 J/Kmol at about T = 40 K, superimposed on the lattice heat, and the anomalous temperature dependence of
quadrupolar interactions. The excited doublet has larger value of $Q_f$ (in the absolute value) so with increasing
temperature quadrupolar interactions observed by means of the Mossbauer spectroscopy should pass a maximum as it was
discussed in Ref. \cite{3}. Such the maximum is rather not expected in case of the $\Gamma _{6}^{1}$ ground state. Such
maxima have been observed in some Yb compounds Ref. \cite{4,5} though, after the conventional interpretation, a strange
interpretation has been proposed.

We do not think that the present set of CEF parameters is the final one. There is a plenty of sets that produce the
shown ground-state eigenfunction (the simplest are those obtained by multiplication of all parameters by a constant
positive value what causes equal spreading of the electronic structure) but surely the got set substantially confines
the searching area of CEF parameters. Although we got this remarkable set of CEF parameters and one of us (RJR) was
advocating by last 12 years for CEF-based model for heavy-fermion phenomena \cite{6,7,8} we would like to express our
big surprise if so well-defined (so extremely thin) atomic-scale magnetism would exist in a metallic compound
YbRh$_{2}$Si$_{2}$, moreover that this compound was regarded as one of the prominent heavy-fermion compound with
itinerant $f$ electrons and with a substantial Kondo temperature (about 25 K). It would be a big surprise for so
adequate description within the CEF approach though we worked hard by last 20 years in the evaluation of CEF and
exchange interactions in different rare-earth compounds, both ionic and intermetallics \cite{9,10}. But it were we who
extended the standard CEF approach to the Quantum Atomistic Solid State (QUASST) Theory \cite{8,11} and recognize that
the standard CEF approach is a giantly correlated electron approach to compounds containing open-shell transition-metal
atoms. In our understanding the Kondo temperature is related to the energy of the first excited CEF doublet and the
Kondo resonance is related to the removal of the Kramers degeneracy that is a source of low-energy, below 0.2 meV,
excitations. These excitations are neutral, spin-like excitations and they are responsible for large low-temperature
specific heat, a hallmark of heavy-fermion physics.

In conclusion, we welcome with the great pleasure the ESR results of Prof. F. Steglich group on heavy fermion metal
YbRh$_{2}$Si$_{2}$. We fully agree with the F. Steglich's interpretation with the localized f electrons existing below
Kondo temperature as it concurs to our long-lasting understanding of heavy-fermion phenomena originating from
Kramers-ion crystal-field states and a deep conviction that the evaluation of CEF interactions is indispensable for
proper description of all transition-metal compounds, including those exhibiting heavy-fermion phenomena and 3$d$ Mott
insulators \cite(12). We have derived CEF parameters of the tetragonal symmetry with a small orthorhombic distortion
that perfectly reproduce the ESR values ($g_\bot$ = 3.561 and $g_\|$ = 0.17). These parameters well reproduce other
thermodynamical properties. The obtained parameters are surely not completely final as searching for CEF parameters is
as a large puzzle (the possible local differentiation of the 4f sites  at low temperatures should be always remembered
- it would lead to site-to-site distribution of $D$; this effect can be discussed as the distribution of Kondo
temperatures) but the discussion of heavy-fermion compounds within the same atomistic approach as conventional
rare-earth compounds is very plausible from the unification point of view. From a methodological point of view - the
electronic structure and the importance of local
distortions can be further experimentally verified. \\
$^{+}$ This set of parameters has been found 30 December 2003 making use of my (RJR) works from 1992-1993. This paper
has been written at the last day of the year 2003. So happy New Year 2004 to everybody. Nice physics to all physicists
in New Year.


\begin{thebibliography}{99}
\footnotesize{
\bibitem{1} J. Sichelschmidt, V.A. Ivanshin, J. Ferstl, C. Geibel and F. Steglich, Phys. Rev. Lett. \textbf{91}, 156401 (2003).
\bibitem{2} J. Custers, P. Gegenwart, C. Geibel, F. Steglich, T. Tayama, O. Trovarelli, N. Harrison,
Acta Phys.Pol. B \textbf{32}, 3211 (2001).
\bibitem{3} R. J. Radwanski, J. Alloys-Compds, \textbf{232}, L5 (1996).
\bibitem{4} P. Bonville, J. Hammann, J.A. Hodges, P. Imbert, G. Jehanno, M.J. Besnus, and A. Meyer, Z. Phys. B Condens. Matter, \textbf 82, 267 (1991).
\bibitem{5} V. Zevin, G. Zwicknagl, and P. Fulde, Phys. Rev. Lett. \textbf{60}, 2331 (1988).
\bibitem{6} R. J. Radwanski, cond-mat/9906287
\bibitem{7} R. J. Radwanski, Report of Center of Solid State Physics CSSP-4/95.
\bibitem{8} R. J. Radwanski, R. Michalski, and Z. Ropka, Acta Phys. Polonica B \textbf{31}, 3079 (2000).
\bibitem{9} R. J. Radwanski, N.H. Kim-Ngan, F.E. Kayzel, J.J.M. Franse, D. Gignoux, D.Schmitt and F.Y. Zhang,
J.Phys.:Condens. Matter \textbf{4}, 8853 (1992).
\bibitem{10} R.J. Radwanski, R. Michalski, Z. Ropka, Physica B \textbf{276-278}, 803 (2000).
\bibitem{11} R. J. Radwanski and Z. Ropka, \textit{Quantum Atomistic Solid State Theory},
cond-mat/0010081.
\bibitem{12}Z. Ropka and R. J. Radwanski, Phys. Rev. B \textbf{67}, 172401 (2003).}
\end{thebibliography}
\end{document}